\documentclass[]{article}
\usepackage{spie}
\usepackage{graphicx}

\def\twotwo{\protect{$\left| 2,2 \right >$}}
\def\oneminusone{\protect{$\left| 1,-1 \right >$}}
\def\twoone{\protect{$\left| 2,1 \right >$}}
\def\onezero{\protect{$\left| 1,0 \right >$}}
\def\twozero{\protect{$\left| 2,0 \right >$}}

\def\Rb87{\protect{${}^{87}\mathrm{Rb}$}}
\def\ie{\emph{i.e.}}
\def\eg{\emph{e.g.}}
\def\etal{\emph{et. al.}}

\title{Recent Experiments with Bose-Condensed Gases at JILA}

\author{D.~S. Hall, J.~R. Ensher, D.~S. Jin,\protect{${}^*$}\ M.~R.
  Matthews, C.~E. Wieman, and E.~A. Cornell\protect{${}^*$}
\skiplinehalf
Joint Institute for Laboratory Astrophysics, National Institute of
  Standards and Technology\\ and University of Colorado, Boulder,
  Colorado 80309-0440\\ and Physics Department, University of Colorado,
  Boulder, Colorado 80309-0440
}

\authorinfo{\protect{${}^*$}Quantum Physics Division, National
  Institute of Standards and Technology}

\begin{document}
\maketitle


\begin{abstract}

  We consider a binary mixture of two overlapping Bose-Einstein
  condensates in two different hyperfine states of \Rb87 with nearly
  identical magnetic moments. Such a system has been simply realized
  through application of radiofrequency and microwave radiation which
  drives a two-photon transition between the two states. The nearly
  identical magnetic moments afford a high degree of spatial overlap,
  permitting a variety of new experiments. We discuss some of the
  conditions under which the magnetic moments are identical, with
  particular emphasis placed on the requirements for a time-averaged
  orbiting potential (TOP) magnetic trap.

\end{abstract}


\section{INTRODUCTION}
\label{sect:intro}
\setcounter{footnote}{1}
The experimental realization of Bose-Einstein condensation (BEC) in the
alkalis\cite{Anderson95,Davis95,Bradley97} has stimulated a flurry of
scientific interest. In addition to single trapped condensates, double
condensates (of two different spin states) have also been
observed.\cite{Myatt97}\footnote{A recent preprint by the MIT
  group\protect{\cite{Kurn97}} also alludes to a multiple-component BEC
  confined in an optical trap.} In the experiments described below, another
double condensate system is described, in which the two condensate
spin states overlap to a higher degree than in the previous
experiment. Moreover, transitions can be driven between the two states
to couple them, produce superposition states, or measure their relative
phase. This double condensate also makes possible a number of novel
experiments which probe the interactions between two interpenetrating
quantum fluids.

This paper describes the new double condensate system. We discuss in
particular the conditions under which spin states are trapped and can
be made to overlap. Since the experiments described here are performed
in a trap with a rotating magnetic field, we also explore the
peculiarities of spectroscopy in such a field.


\section{CONDENSATE PRODUCTION}
\label{sect:condprod}

We use a new, third-generation JILA BEC apparatus, in which the
time-averaged orbiting potential (TOP) magnetic trap\cite{Petrich95}
of generation I\cite{Anderson95} is combined with the double
magneto-optical trap (MOT) system\cite{Myatt96} of generation
II\cite{Myatt97} to produce condensates containing up to one million
atoms. \Rb87 atoms are first collected in a vapor cell
MOT\cite{Raab87,Monroe90} and then magnetically guided through a
transfer tube into a second ultrahigh vacuum ($10^{-12}$~Torr)
MOT\cite{Myatt96}, where up to $10^9$ atoms are collected in
$\sim15$~seconds.  After further cooling in optical molasses and
optically pumping to the $\left| F=1,m_f = -1 \right >$ hyperfine
state, the trapped atoms are loaded into a magnetic trap. The atoms
are further cooled by forced evaporation,\cite{Hess87} in which an
applied radiofrequency (rf) magnetic field drives transitions in the
most energetic atoms to untrapped states (\eg, \onezero).  We adjust
the final evaporation frequency to choose the fraction of atoms which
are condensed into the ground state,\cite{Anderson95} which can range
from 0\%, a cold thermal cloud, to nearly 100\%, a pure condensate.
The atoms are held in the trap for up to several seconds and
subsequently released and allowed to expand ballistically for
imaging.\cite{Anderson95} The entire cycle typically takes under one
minute, and the apparatus is reliable enough to operate unattended,
producing nearly identical condensates without interruption for over an
hour at a time.

The technique of resonant absorption imaging\cite{Anderson95} is used
to probe the cloud. In this destructive process, a brief pulse of
repump light ($5S_{1/2}, F=1$ to $5P_{3/2}, F'=2$) transfers the
\oneminusone\ atoms to the $F=2$ hyperfine level. The atoms are
subsequently illuminated by the probe beam, which drives the
$5S_{1/2}, F=2$ to $5P_{3/2}, F'=3$ cycling transition. Atoms scatter
the light out of the probe beam and the resulting shadow is imaged
upon a charge-coupled device (CCD) array.  We process these data to
extract the optical depth as a function of position, which
permits us to determine the size of the cloud and number of atoms of
which it is composed, as well as thermodynamic quantities such as its
temperature $T$.\cite{Ensher96}

\section{TWO OVERLAPPING CONDENSATES}

In a recent experiment at JILA, Myatt and co-workers\cite{Myatt97}
produced two overlapping condensates in the \twotwo\ and \oneminusone\
hyperfine states of \Rb87. Due to their different magnetic moments,
their mutual repulsion, and the effect of gravity, the condensates occupied
slightly different regions in the trap and could not be made to
overlap one another completely. A third spin state of \Rb87\ may also be
trapped which has (to first order) the same magnetic moment as the
\oneminusone\ state, and (as will be shown below) can be made to
overlap more completely with it. This system permits the realization
of a rudimentary Rb clock in a magnetic trap, and provides a
vehicle for understanding the interplay and dynamics between two
condensates trapped in different spin states without the
complications introduced by the lack of condensate overlap.

\subsection{Magnetic Moments}
\label{sec:magmom}

\begin{figure}
\begin{center}
\mbox{\includegraphics*[height=7cm]{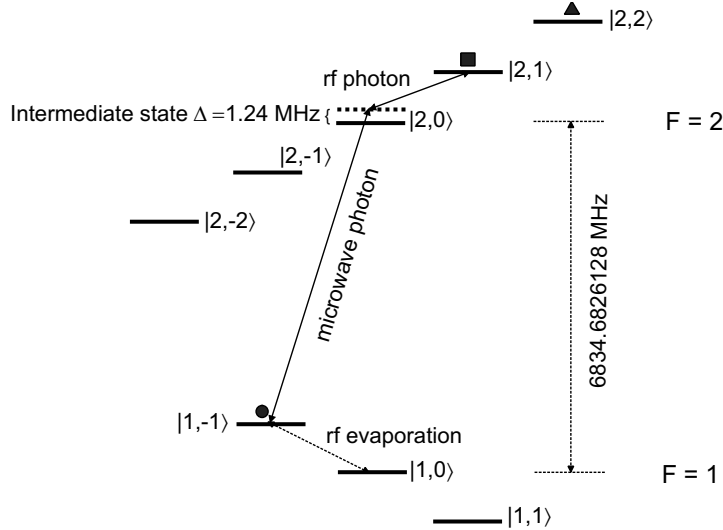}}
\end{center}
\caption[states]{\label{fig:states}
  A schematic of the \Rb87\ ground states. Atoms may be magnetically trapped
  in the states indicated by the solid polygons. We begin with atoms
  in the \protect{$\left|F=1,m_f=-1\right>$} state; evaporative
  cooling proceeds by driving rf transitions to the untrapped
  \onezero\ state. The two-photon transition to the trapped \twoone\
  state is accomplished as shown, with an intermediate state detuning
  of 1.24~MHz.}
\end{figure}

In the low-field limit, the magnetic energy of a state characterized
by magnetic moment $\mathbf{\mu}$ and quantum number $m_f$ quantized
along the direction of the field $B$ is
\begin{equation}
\label{eqn:magnrg}
U = -{\mathbf{\mu \cdot B}} = g_F m_f \mu_B B
\end{equation}
where $g_F$ is the Land\'e $g$-factor and $\mu_B$ is the Bohr
magneton. The magnetic moment precesses about the field at the Larmor
frequency
\begin{equation}
\omega_L = -\frac{g_F \mu_B B}{\hbar}.
\label{eqn:larmor}
\end{equation}
States for which the product $g_F m_f$ is positive minimize their
energy in low fields (``weak-field seekers'') and may be trapped
in the minimum of a magnetic field. For ${}^{87}\mathrm{Rb}$, this
leads to three trappable states (Fig.~\ref{fig:states}): the
\twotwo\ and \twoone\ states in the $F=2$ hyperfine manifold
(where $g_F=+\frac{1}{2}$), and the \oneminusone\ state in the
$F=1$ manifold (where $g_F=-\frac{1}{2}$). Note that, because of
the sign of $g_F$, the trapped \oneminusone\ state precesses in a
sense opposite to that of the other two trapped states. This
difference gives rise to some subtle behavior in the rotating
magnetic field of the TOP trap, as we shall see below.

In the absence of gravity, all three of the trapped states sit atop
one another, although the \twotwo\ atoms are bound a factor of two
more tightly by their larger magnetic moment. The centers of the
atomic clouds are displaced (``sag'') in the presence of gravity by an
amount which scales as $g/\mu$, where $g$ is the acceleration due to
gravity. As a result, the \twotwo\ atoms sag about half as much than
their \twoone\ and \oneminusone\ counterparts, and the degree of
spatial overlap between the states is reduced.  To a first
approximation, however, the \twoone\ and \oneminusone\ atoms sag
equally in the trap; consequently, we choose these two states for the
fully overlapping condensate experiments.

\begin{figure}[p]
\begin{center}
\mbox{\includegraphics*[width=0.9\linewidth]{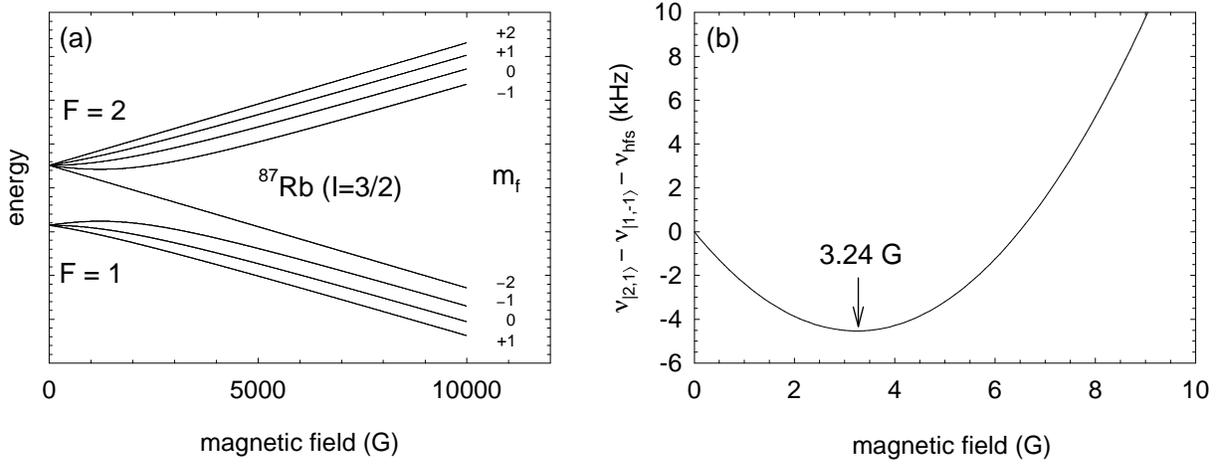}}
\end{center}
\caption[ediff]{\label{fig:ediff}
  (a) The shifts of the \Rb87\ energy levels as a function of the
  magnetic field, as calculated from the Breit-Rabi formula
  (Eq.~\protect\ref{eqn:breitrabi}). (b) The frequency difference
  between the \twoone\ and \oneminusone\ states (less the zero-field
  hyperfine splitting) is shown for weak magnetic fields. When the
  slope of the curve is zero (\protect{$B=3.24$~G}) the two states
  have the same effective magnetic moment.}
\end{figure}

Beyond first order, the magnetic moments of the \twoone\ and
\oneminusone\ states depend (differently) upon the applied magnetic
field, and a more exact analysis is required. The Hamiltonian for
these atoms includes couplings between the nuclear spin $I=3/2$, the
electronic spin $J=1/2$, and the externally applied magnetic field; this
is expressed\cite{Corney88}
\begin{equation}
  \hat{H}_{\mathrm{BR}} = \frac{h
    \nu_{\mathrm{hfs}}}{2}{\mathbf{I \cdot J}} + g_J\mu_B{\mathbf{J
      \cdot B}} - g_I\mu_n{\mathbf{I \cdot B}},
\label{eqn:brhamiltonian}
\end{equation}
where $\nu_{\mathrm{hfs}}=6834682612.8$~Hz is the zero-field hyperfine
splitting in \Rb87,\cite{arimondo77} $g_J$ and $g_I$ are the
electronic and nuclear $g$-factors, and $\mu_n$ is the nuclear magneton. The
energy of a particular state as a function of the magnitude of the
magnetic field is found by diagonalizing
Eq.~(\ref{eqn:brhamiltonian}), which yields the well-known Breit-Rabi
formula:\cite{Corney88}
\begin{equation}
  E_{\mathrm{BR}}(B) = - \frac{h \nu_{\mathrm{hfs}}}{2(2I + 1)}-
  g_I\mu_nBm_f \pm \frac{h \nu_{\mathrm{hfs}}}{2}\sqrt{1 + \frac{4 m_f
      x}{2I + 1} + x^2}
\label{eqn:breitrabi}
\end{equation}
where the upper (lower) sign is taken for $F=2$ ($F=1$), and the
parameter $x$ is defined as
\begin{equation}
x = \frac{(g_J \mu_B + g_I \mu_n)B}{h\nu_{\mathrm{hfs}}}.
\end{equation}
A plot of the splitting between the \twoone\ and \oneminusone\ states
is shown in Fig.~\ref{fig:ediff}(b). The magnetic moments of the two
states are identical when the slope of this difference becomes zero,
\ie, at $B=3.24$~G. At this field\footnote{This requirement is
  only on the magnitude of the field seen by the atoms, and can arise
  from a combination of (rotating) bias fields and quadrupole
  components.} two noninteracting atomic clouds (or condensates) will
overlap.

\subsection{Creating the Double Condensate}

The double condensate is produced by driving atoms from the
\oneminusone\ condensate to the \twoone\ state with a two-photon
excitation\cite{Newbury95} involving a microwave photon at $\sim
6.8$~GHz and an rf photon at $\sim 1$~MHz.\footnote{Note that this
  scheme differs from that used in the previous JILA
  experiment,\protect\cite{Myatt97} in which the two condensates were
  produced simultaneously during the evaporative cooling cycle.}
The two-photon drive is detuned 1.24~MHz from the intermediate
\twozero\ state, as shown in Fig.~\ref{fig:states}. The rf pulse is
generated by a Wavetek~395 synthesizer and coupled into the system in
the same manner as the rf used for forced evaporative cooling.  The
microwave radiation is generated by an HP~8672A frequency synthesizer,
amplified to approximately 1~W by a travelling wave tube amplifier, and
coupled into the system through a truncated waveguide. The coupling is
rather inefficient, and the resulting two-photon Rabi frequency is
only 1.1~kHz. For stability and absolute hyperfine interval
measurements (described below) both synthesizers are frequency-locked
to the 10~MHz reference of an HP~53570A global positioning system
(GPS)-stabilized time and frequency reference receiver.

\begin{figure}
\begin{center}
\mbox{\includegraphics*[height=7cm]{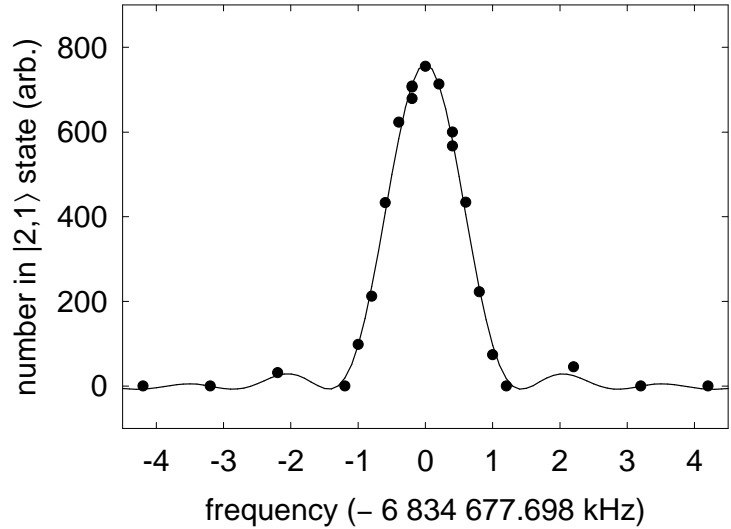}}
\end{center}
\caption[sinc]{\label{fig:sinc}
  Absorption line for the two-photon excitation between the
  \oneminusone\ and \twoone\ states, measured after the atoms were
  released from the trap. The width of the line is determined by the
  length of the pulse. Similar lines were also measured in the trap.}
\end{figure}

By varying the rf detuning we can sweep the drive across the
transition frequency and measure the number of atoms in the \twoone\
condensate (Fig.~\ref{fig:sinc}).  The repump light is blocked in the
imaging sequence in order to prevent atoms in the \oneminusone\ state
from being imaged. Each point corresponds to a measurement with a pure
condensate (\ie, evaporated to the point at which there was no visible
thermal cloud).  For this particular measurement the transfer pulse
was applied after the atoms were released from the trap, but identical
results are obtained with atoms transferred to the upper state while
in the trap. The width of the line is consistent with the length of
the applied pulse, which is approximately $500~\mu$sec.

Strictly speaking, any transfer of atoms to the upper state which is
incomplete results in a single condensate in a superposition of the
two states. The probe light projects these states after the condensate
is released from the trap. Unless the relative phase between the two
condensates is to be measured (as in the Ramsey method described
below) there is little experimental difference between the
superposition state and two independent condensates.

\subsection{Effect of a Rotating Field}

The preceding analysis is sufficient for a static magnetic trap, such
as that used in the experiment of Myatt \etal.\cite{Myatt97}
Additional complications arise, however, when a time-dependent
magnetic trap (such as the TOP trap) is used to provide the
confinement. The TOP trap involves a magnetic field of magnitude
$B_b$ rotating counterclockwise (as viewed from the
positive $z$-axis) at angular frequency $\omega_t$ in the $xy$-plane,
\begin{equation}
  {\mathbf{B_b}}(t)=B_b(\cos\omega_t t
  {\mathbf{\hat{x}}} + \sin\omega_t t {\mathbf{\hat{y}}}).
\end{equation}
The effective Hamiltonian in a frame co-rotating with the magnetic
field transforms as
\begin{equation}
  \hat{H}_{\mathrm{eff}} = R(-\omega_t t) \hat{H}_{\mathrm{BR}}
  R^\dag(-\omega_t t) - i \hbar R^\dag(-\omega_t
  t)\frac{\partial}{\partial t} R(-\omega_t t)
\end{equation}
with the time-dependent rotation operator $R$ defined by
\begin{equation}
R(-\omega_t t) = \exp({\frac{i}{\hbar}F_z \omega_t t}).
\end{equation}
The operator $F_z$ is the $z$-component of the total (nuclear plus
electronic) spin vector, and the sign of the argument of $R$ is chosen
to rotate the coordinate axes in the same sense as the field. The
Breit-Rabi Hamiltonian $\hat{H}_{\mathrm{BR}}$ is invariant under the
transformation. We therefore obtain
\begin{equation}
\hat{H}_{\mathrm{eff}}=\hat{H}_{\mathrm{BR}} -
F_z\omega_t.
\end{equation}
For weak fields (\ie, for $\omega_L \gg \omega_t$) this additional term
may be approximated as an effective magnetic field $B_\omega$ pointing
along the $z$-axis which causes Larmor precession of the spin at the
trap rotation frequency, \ie,
\begin{equation}
  \omega_t = -\frac{g_F \mu_B B_\omega}{\hbar} \Longrightarrow
  B_\omega = - \frac{\hbar \omega_t}{g_F \mu_B}.
\end{equation}
The total spin vector for each state precesses about the magnetic
field in a direction specified by the sign of its Land{\'e}
$g$-factor $g_F$; as a consequence, the direction of the effective
magnetic field $B_\omega$ is also determined by this sign. For our
two trapped states we can take out the overall sign of $B_\omega$
explicitly and write
\begin{equation}
B_{\mathrm{total}}=\sqrt{B_b^2 + (\mp |B_\omega|)^2}
\label{eqn:effB1}
\end{equation}
where the upper (lower) sign is for the \twoone\ (\oneminusone) state.

Up to this point we can see that the effective field is the same for
both states. In TOP traps, however, there is also a magnetic
quadrupole field
\begin{equation}
  {\mathbf{B_q}}= B'_q(x{\mathbf{\hat{x}}} + y{\mathbf{\hat{y}}} -
  2z{\mathbf{\hat{z}}})
\end{equation}
(where $B'_q = \partial B / \partial x$ and the sign of the quadrupole
has been chosen such that the field points toward the center of the
trap on the $z$-axis). Gravity causes the equilibrium displacement of
the atoms to sag below the center of the trap as defined by the
quadrupole gradient. If the force on the atoms due to gravity is in
the $-z$-direction and the atoms are displaced some distance $z$ from
the center of the trap, Eq.~\ref{eqn:effB1} must be modified to read
\begin{equation}
B_{\mathrm{total}}=\sqrt{B_b^2 + (-2 B'_q z \mp |B_\omega|)^2}
\label{eqn:effB2}
\end{equation}
where the upper (lower) sign is taken for atoms in the \twoone\
(\oneminusone) state. As a first result, we note that the measured
transition frequency between the two states in the TOP trap will not
be that shown in Fig.~\ref{fig:ediff}(b), which assumes that both
states see the same field. In the TOP trap, each state sees a field
different from that seen by the other since the rotating term
$B_\omega$ adds to the field produced by the quadrupole gradient for
one state and is subtracted from it for the other.

\begin{figure}[p]
\begin{center}
\mbox{\includegraphics*[height=7cm]{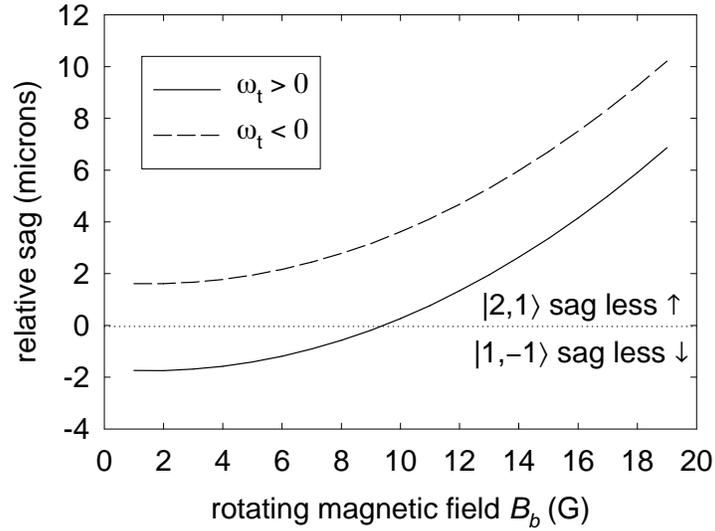}}
\end{center}
\caption[sag]{\label{fig:sag}
  Calculated relative displacement between the two states as a function of
  magnetic field for \protect{$\omega_t=2\pi\cdot 7202$}~Hz and a
  quadrupole gradient \protect{$B'_q = 61.5$}~G/cm.  For
  \protect{$\omega_t > 0$} the TOP trap bias field rotates in the
  counterclockwise direction as viewed from the {\protect$z$}-axis.}
\end{figure}

The equilibrium position of the atoms is where the total force
on the atoms is zero, or
\begin{equation}
\frac{\partial}{\partial z}\left [E_{\mathrm{BR}}(B_{\mathrm{total}})\right]
= M_{\mathrm{Rb}}g.
\end{equation}
Our second result is that the two equilibrium displacements will in
general not be the same when $B_b=3.24$~G. We may, however, find new
conditions under which the two states once again share an equilibrium
displacement in the TOP trap by calculating the relative sag as a
function of the magnetic quadrupole gradient as well as the bias field
magnitude, rotation frequency, and sense of rotation (all of which are TOP trap
parameters which we can adjust within certain limits).  For instance,
Fig.~\ref{fig:sag} shows the calculated sag for a trap rotation
frequency $\omega_t = 2 \pi \cdot 7.202$~kHz and a quadrupole gradient
$B'_q = 61.5$~G/cm, where the upper curve is for clockwise rotation
($\omega_t < 0$) and the lower curve for counterclockwise rotation
($\omega_t > 0$). In this case, the relative sag is calculated to be
zero when $B=9.4$~G and the field rotates in the counterclockwise
sense. Note there is no value of $B_b$ for which the equilibrium
displacements in the trap are the same for atoms in both states when
the TOP field rotates in the clockwise sense.

\begin{figure}
\begin{center}
\mbox{\includegraphics*[width=0.9\linewidth]{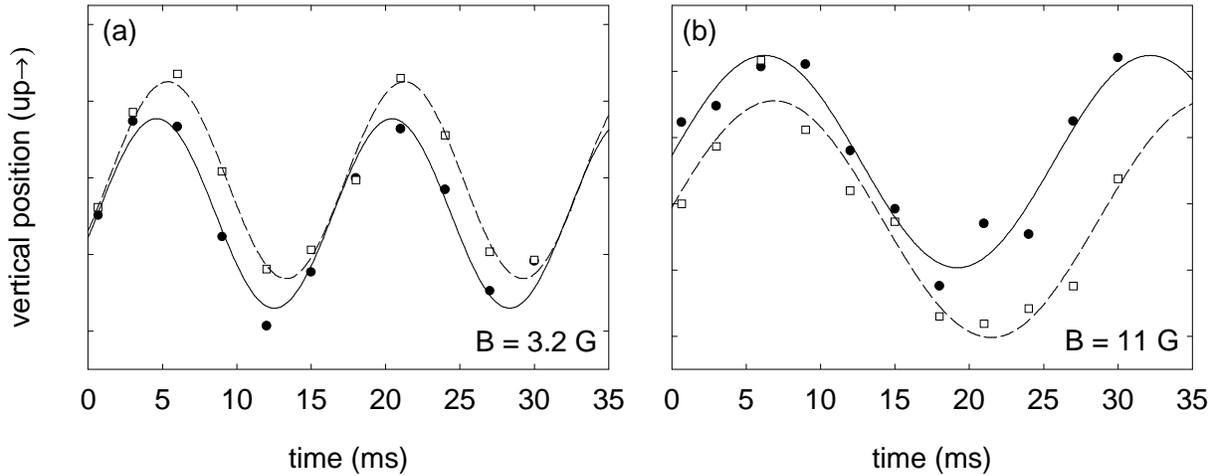}}
\end{center}
\caption[osc]{\label{fig:osc}
  The relative centers of oscillation of the two states can be seen to
  differ in this set of plots which show the center of mass
  (``sloshing'') motion in the vertical direction. At 3.2~G (a), the
  \twoone\ atoms (solid circles) are seen to sag less than the
  \oneminusone\ atoms (hollow squares); the situation is reversed at
  11~G (b). These observations agree with the theoretical prediction
  of Fig.~\protect\ref{fig:sag}. (The lines are fit to the data.)}
\end{figure}

In order to confirm the relative sag as predicted by the theory for
the rotating magnetic field, we drove the $m=0$ center-of-mass
(``sloshing'') motion by discontinuously changing the trap fields
(and, hence, oscillation frequencies) for pure \oneminusone\ and
\twoone\ condensates. The resulting condensate axial motion is shown in
Fig.~\ref{fig:osc}. For a field of 3.2~G, the \twoone\ atoms are seen
to oscillate about an equilibrium position somewhat higher than that
of the \oneminusone\ atoms; the reverse is true for 11~G. These
observations agree with the theoretical prediction of
Fig.~\ref{fig:sag}.

\section{APPLICATIONS, PRELIMINARY RESULTS, AND CONCLUSIONS}

The \onezero$\rightarrow$ \twozero\ ``clock'' transition in \Rb87\
is widely used as a frequency reference. In addition to systematic
effects, the precision to which such an atomic transition can be
measured is limited by the time over which the transition can be
observed. By using condensates confined in a magnetic trap, the
observation time can be on the order of the lifetime of the
condensates, which may be up to several seconds.  Unfortunately, the
states traditionally used in the ``clock'' transition cannot be
magnetically trapped since $m_f = 0$.

\begin{figure}
\begin{center}
\mbox{\includegraphics*[height=7cm]{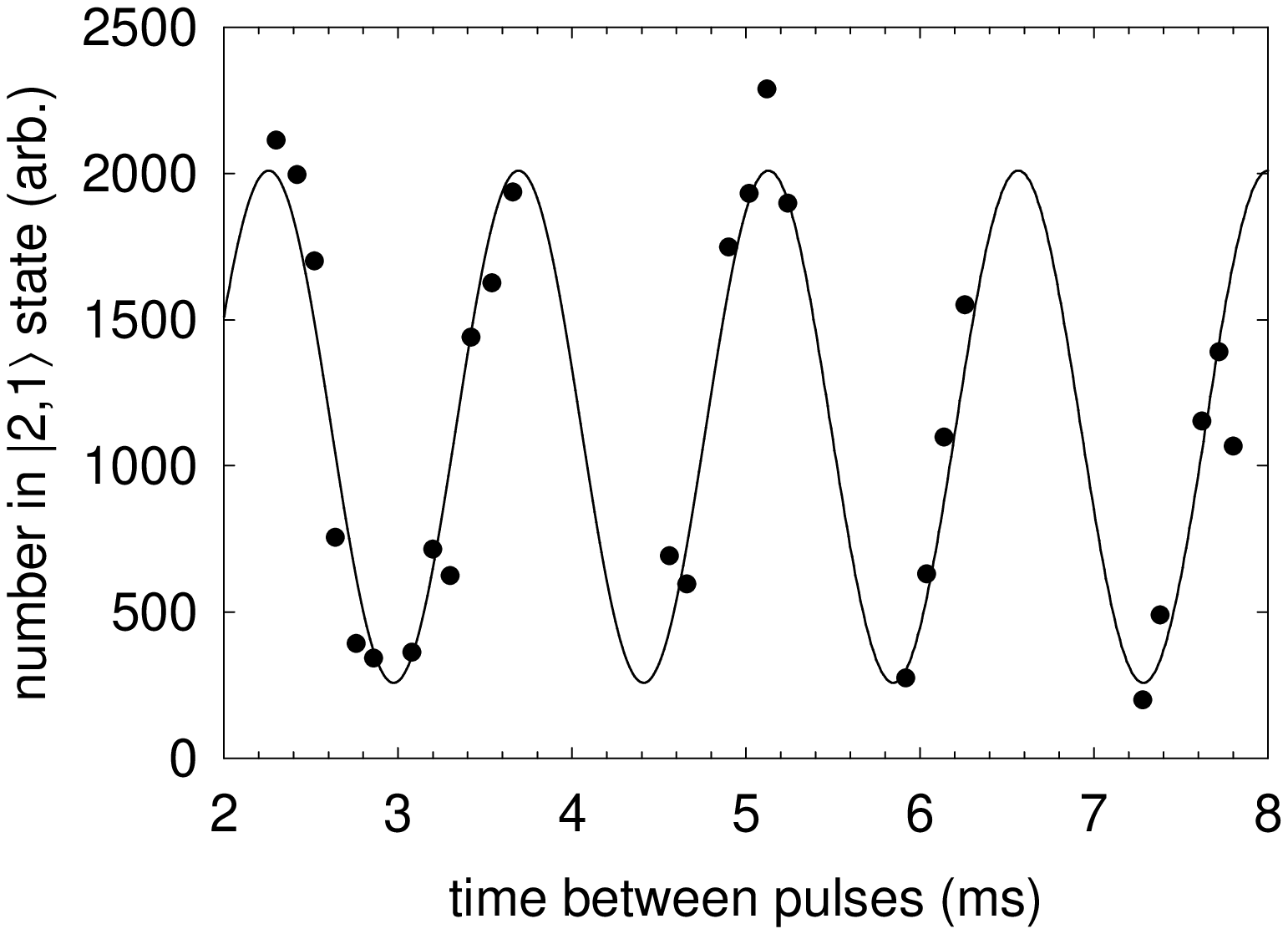}}
\end{center}
\caption[fringe]{\label{fig:fringe}
  Ramsey fringes, measured using two time-separated pulses of rf and
  microwave radiation on condensate atoms released from the trap and
  allowed to expand ballistically. The frequency measured is the
  difference between the GPS-stabilized local oscillator frequency and
  the frequency of the atomic hyperfine transition. Similar fringes
  have now been observed with magnetically trapped atoms.}
\end{figure}

As noted in Sec.~\ref{sec:magmom} (above), the \twoone\ to \oneminusone\
transition frequency is (like the clock transition) independent of $B$
to first order. We can measure this frequency by using Ramsey's
technique of separated oscillatory fields (SOF).\cite{Ramsey56} Unlike
a beam experiment, the fields here occur as two pulses separated in
time rather than two interaction regions separated in space. We begin
with atoms in the \oneminusone\ state. A first $\pi / 2$ pulse creates
a condensate in a superposition of the two states, at which point its
wavefunction freely evolves at the frequency of the hyperfine
splitting $\omega_{\mathrm{clock}}=\omega_{\left |2,1\right
  >}-\omega_{\left | 1,-1 \right >}$. Our local oscillator (the sum of
the microwave and rf frequencies) evolves at $\omega_{\mathrm{LO}}$.
The number of atoms observed in the upper state after some interaction
time $t$ is given by
\begin{equation}
  \frac{N_{2,1}}{N_{\mathrm{tot}}}=\sin^2\left((\omega_{\mathrm{clock}}-
    \omega_{\mathrm{LO}})\frac{t}{2} + \phi_0 \right )
\end{equation}
where $\phi_0$ is an unimportant relative phase originating from the
details of the pulse timing. We detune the local oscillator roughly
1~kHz from the transition and measure the number of atoms in the
\twoone\ condensate for various interaction times. From a fit to
the data we obtain a measurement of the detuning, as shown in
Fig.~\ref{fig:fringe}.\footnote{This single measurement does not give
  us the \emph{sign} of the detuning, and we are required to choose a
  different detuning and repeat the measurement to get at this
  quantity.}

If the equilibrium positions of the two condensates are not identical
then the wavefunctions for the two condensate states will begin to
separate in space and will not effectively interfere with one another
when the second $\pi/2$ pulse is applied.  The data presented in this
figure determine the hyperfine splitting between the two states to
about 1 part in $10^9$, and were taken with atoms released from the
trap\cite{Monroe91} and allowed to expand ballistically in only the
rotating magnetic field.  Ramsey fringes have recently been obtained
with condensates confined in the magnetic trap (after having adjusted
the trap parameters to make the condensate equilibrium positions
identical) and will be considered in a future paper.

Another set of future experiments focuses upon the interactions
between two condensates in two different hyperfine states, which
constitutes one of a class of studies of mixed
condensates.\cite{Ho96,Esry96} Choosing the equilibrium displacements
to be identical simplifies the interpretation of the dynamical
behavior of the two condensates, and maximizes the overlap which can
be achieved. In particular, the system described here permits very
good control over the relative number of atoms in the two states.
Experiments under consideration include the measurement of the
relative scattering length of the two states by examining the mode
spectrum excited by transferring all of the atoms from the
\oneminusone\ state to the \twoone\ state, and the phase separation
dynamics of the two condensates.

In conclusion, we have produced a new two-condensate system out of the
\oneminusone\ and \twoone\ hyperfine states of \Rb87. Due to their
very similar magnetic moments these two condensates can be made to
overlap to a very high degree in both static and rotating-field
magnetic traps. This system, and others like it, may be used as
a vehicle for studying the dynamics of interpenetrating quantum fluids
and for precision metrology.


\acknowledgments

This work is supported by the National Institute of Standards and
Technology, the National Science Foundation, and the Office of Naval
Research.  The authors would also like to thank the other members of
the JILA BEC Collaboration, and in particular J.~L. Bohn, for their
contributions and discussions on these topics.



This paper first appeared in the Proceedings of the SPIE, Volume
3270, Pages 98--106 (1998). Equations (2) and (10), and their
accompanying text, have been corrected for this postprint edition.

\end{document}